\documentclass[10pt]{iopart}

\usepackage{iopams}
\usepackage{siunitx}
\usepackage{graphicx}
\usepackage{todonotes}
\usepackage{subcaption}
\usepackage{hyperref}

\DeclareSIUnit\townsend{Td}

\begin{document}
\title[Effects of gas temperature on Monte-Carlo simulations]{Effects of gas temperature on Monte-Carlo simulations of charged particles drift in gaseous medium}

\author{
	Michele Renda$^1$, 
	Iulia Stefania Trandafir$^{1,2}$
}

\address{$^1$ IFIN-HH, Particles Physics Department, M\u{a}gurele, Romania}
\address{$^2$ Faculty of Physics, University of Bucharest - M\u{a}gurele, Romania}

\ead{michele.renda@cern.ch}

\begin{abstract}
	We present the derivation of kinetic formulas modeling the microscopic interaction of a charged particle withing a molecular gas under effect of thermal motion. Both elastic and inelastic processes are taken in account. The results were verified to reproduce the non-thermal formulas when the target molecule velocity is set to zero. A set of simulation is provided to highlight the effects in Argon and 
	Carbon tetrafluoride. Our results can applied in Monte-Carlo simulation of particle drift at energies of the same order of the thermal kinetic energy of the buffer gas.
\end{abstract}

\submitto{\JPD}
\maketitle


\section{Introduction}

Monte-Carlo simulations have an important role in the calculation of the swarm coefficients of electrons and ions drifting in gaseous medium under the effect of an electromagnetic field. Historically, such calculations were performed thus the resolution of the Boltzmann transport equation \cite{boltzmann1896vorlesungen,ness1994multi} and this method is still used today when the predominant class of collisions is composed by elastic processes.

However, thanks to the advance of computing technology, it is possible to solve these problems using methods which perform a microscopic simulation of a large number of interactions and then perform a statistical analysis to extract macroscopic attributes of the system. The main advantage of this approach is that we can better simulate the inelastic processes, such as ionization and excitation, being only limited by the computing power at our disposal.

Several software tools were developed to be able to simulate drift of charged particles, such as \texttt{Magboltz} \cite{biagi1999monte} and \texttt{METHES} \cite{rabie2016methes} (and their respective Cython and Python porting, \texttt{PyBoltz} \cite{atoum2020electron} and \texttt{pyMETHES} \cite{pymethes2020chachereau}). Such tools are a valuable resource for the low pressure plasma community, being used for the simulation of gaseous detector used in high-energy physics experiments.

The authors of this article were involved also on the development of a software tools, \texttt{Betaboltz} \cite{renda2020betaboltz}, which uses some formulas discusses in this article to perform microscopic simulation of charged particle in gaseous medium.

The simulation of microscopic collisions is not a complex processes and can be divided in four elementary steps:
\begin{enumerate}
	\item Calculation of the free path for each charged particle.
	\item Chose of the interaction process (elastic, inelastic, etc.).
	\item Determination of the new particle direction.
	\item Determination of the energy transfer.
\end{enumerate}

Because these steps must be repeated thousands of times, it is critical that they should be implemented as efficient as possible. The first step can be performed using the \textit{null-collision} technique as described by Skullerud \cite{skullerud1968stochastic} and improved by Lin and Bardsley \cite{lin1977monte}, Brennan \cite{brennan1991optimization} and Koura \cite{koura1998improved}.

Once the free path was chosen for the current particle, and confirmed it is a \textit{not null} collision, we have to select the physics process that will take place. As shown by Fraser \cite{fraser1987monte}, we can perform a random selection, weighted on the process cross-section values seen by the particle at its current energy.

The last step consist on calculate the new direction of the particle and energy transfer. To proper choose a deflection angle, we can assume inelastic collision to be isotopic, while elastic collisions require the knowledge of the differential cross-section of the process. Unfortunately, such tables are quite rare and limited to a few common gases. However, it is possible to use integral cross-section tables, which are widely available in literature, to generate pseudo-differential cross-section tables using the methods presented by Okhrimovskyy et al. \cite{okhrimovskyy2002electron} or by Longo and Capitelli \cite{longo1994simple}.

To calculate the energy transfer, we can use the formula presented by Fraser \cite{fraser1987monte}, which provides, for inelastic collisions, the following relation:
\begin{eqnarray}
\fl \Delta E_1 =& \frac{m_2}{(m_1+m_2)} \varepsilon_k / E_1 - \frac{2 m_1 m_2}{(m_1 + m_2)^{2}} \label{eq:delta_e_ine} \\
& \left\{\left[1-\frac{(m_1 + m_2)}{m_2} \varepsilon_k / E_{1}\right]^{1 / 2} \cos \theta_1-1\right\} \nonumber 
\end{eqnarray}
which, for elastic collisions, reduces to:
\begin{equation}
\Delta E_1 = \frac{2 m_1 m_2 (1-\cos \theta_1)}{(m_1 + m_2)^{2}} \label{eq:delta_e_ela}
\end{equation}

Here, $m_1$ is the mass of the charged particle, an electron or ion, named \textit{bullet}, $m_2$ is the mass of the gas molecule, from now on label as \textit{target}, $E_1$ is the bullet energy, $\theta_1$ is the deflection angle in center of momentum frame (more on this in next section) and $\varepsilon_k$ the threshold energy of the inelastic process.

Analyzing \eref{eq:delta_e_ine} and \eref{eq:delta_e_ela}, we may notice there is no reference about the target molecule energy but only to its mass. The reason is that the target molecule is considered at rest. This approximation holds well for the energy domain in common particle drift experiments. However, when the drift fields goes below $\approx$ \SI{0.01}{\townsend}, the mean energy for an electron became comparable to the thermal energies of the gas molecules ($\approx$ \SI{25}{\milli\electronvolt} at standard conditions).

In this article, we will derive and discuss how \eref{eq:delta_e_ine} and \eref{eq:delta_e_ela} were derived, and we will provide a set of equations which can be used to get a more precise simulation at low reduced fields.

\section{Conventions and reference frames}
When discussing microscopic collisions between a charged particle and a molecule, it is important to define the right frame of reference. In an experimental setup, it is commonly used the laboratory frame of reference. However, when handling collisions between two moving objects, it is easier to work in the center-of-momentum reference frame. Indeed, we can define, three reference frames:
\begin{enumerate}
	\item Global laboratory reference frame
	\item Local laboratory reference frame
	\item Center-of-momentum reference frame
\end{enumerate} 

\begin{figure}
	\centering
	\includegraphics[width=\linewidth]{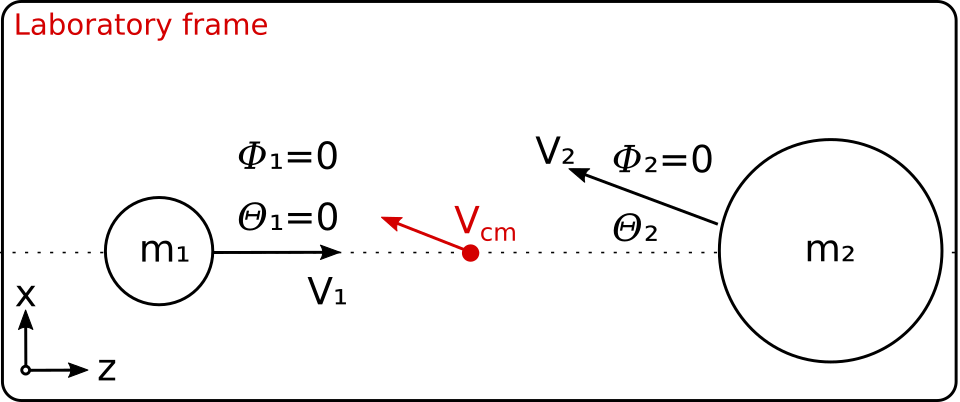}
	\caption[Laboratory reference frame]{In this frame, the $z$-axis is aligned along $\textbf{V}_1$, while $\textbf{V}_2$ falls in the $xz$-plane. $\textbf{V}_{cm}$ is not null and can be calculated using \eref{eq:com}.}
	\label{fig:lab_frame}
\end{figure}

\begin{figure}
	\centering
	\includegraphics[width=\linewidth]{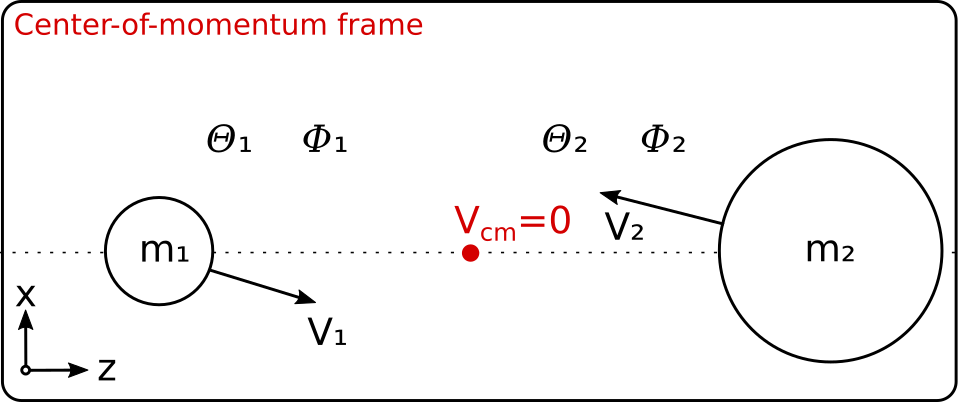}
	\caption[Center-of-momentum reference frame]{In this frame, $\textbf{V}_{cm}$ is null and will remain so before and after the collision.}
	\label{fig:com_frame}
\end{figure}

The first one, is the normal reference frame which is arbitrary aligned, usually with one of its axis is a relevant axis of the experimental setup. The second one, is just a rotation of the first reference frame, such as the $z$-axis will be aligned along the velocity of the bullet particle. The last one, will be relative to the center-of-momentum frame, defined as:
\begin{equation}
	\textbf{V}_{cm} = \frac{m_1 \textbf{V}_1 + m_2 \textbf{V}_2}{m_1 + m_2} \label{eq:com}
\end{equation}
where, $m_1$ and $m_2$ are the bullet and target mass, and $\textbf{V}_1$, $\textbf{V}_2$ their velocities. We think it is important to remark that, while the global laboratory reference frame is static during the simulation, the other two frames are different for each collision. In this article we will ignore the global laboratory frame, and we will focus only on the local laboratory and the center-of-momentum frame (shown respectively in \fref{fig:lab_frame} and \ref{fig:com_frame}).

In the remaining part of the article, we will use these conventions:
\begin{itemize}
	\item Capital symbols will be used for laboratory frame quantities
	\item Lowercase symbols will be used for center-of-momentum frame quantities
	\item Bold symbols refer to Cartesian vectors
	\item The lower script 1 will be used for the bullet, the electron or ion which is drifter by the electromagnetic field.
	\item The lower script 2 will be used for the target, gas molecules which may have thermal kinetic energies.
	\item The apostrophe will be used to mark \textit{after-collisions} quantities, while plain symbols will be used for \textit{before-collsion} quantities or constant attributes such as masses.
\end{itemize}

\section{Particle collisions in center of momentum frame}

In the center-of-momentum frame, we can threat the collisions with the common conservation laws, keeping in mind that, due to how this frame is defined, the total moment of the system is zero:
\begin{eqnarray}
m_1 \textbf{v}_1  + m_2 \textbf{v}_2  = 0 \label{eq:mom_cons_1} \\
m_1 \textbf{v}_1' + m_2 \textbf{v}_2' = 0 \label{eq:mom_cons_2}\\
\frac{1}{2} m_1 (v_1)^2 + \frac{1}{2} m_2 (v_2)^2 = \frac{1}{2} m_1 (v_1')^2 + \frac{1}{2} m_2 (v_2')^2 + \varepsilon_k \label{eq:ene_cons}
\end{eqnarray}

Combining \eref{eq:mom_cons_1} and \eref{eq:mom_cons_2}, we get the relations:
\begin{equation}
\textbf{v}_2  = -\frac{m_1}{m_2} \textbf{v}_1   \\
\textbf{v}_2' = -\frac{m_1}{m_2} \textbf{v}_1' \\
\end{equation}
and using \eref{eq:ene_cons}:
\begin{eqnarray}
v_1' = \sqrt{v^2_1 - \frac{2 m_2}{m^2_1 + m_1 m_2} \varepsilon_k} \label{eq:full_v1} \\
v_2' = \sqrt{v^2_2 - \frac{2 m_1}{m^2_2 + m_1 m_2} \varepsilon_k} \label{eq:full_v2} \\
\varepsilon_1' = \varepsilon_1 - \frac{m_2}{m_1 + m_2} \varepsilon_k \label{eq:full_e1} \\
\varepsilon_2' = \varepsilon_2 - \frac{m_1}{m_1 + m_2} \varepsilon_k \label{eq:full_e2} 
\end{eqnarray}

\section{Energy transfer in laboratory frame}

Moving from the laboratory frame to the center-of-momentum frame, can be done using the relations:
\begin{eqnarray}
\textbf{v}_1 &= \textbf{V}_1 - \textbf{V}_{cm} \label{eq:llf_to_com_1} \\
\textbf{v}_2 &= \textbf{V}_2 - \textbf{V}_{cm} \label{eq:llf_to_com_2} 
\end{eqnarray}
and vice-versa:
\begin{eqnarray}
\textbf{V}_1' &= \textbf{V}_{cm} + \textbf{v}_1' \label{eq:com_to_llf_1} \\
\textbf{V}_2' &= \textbf{V}_{cm} + \textbf{v}_2' \label{eq:com_to_llf_2} 
\end{eqnarray}

To get the final velocities, we can replace \eref{eq:full_v1}, \eref{eq:full_v2} in \eref{eq:com_to_llf_1} and \eref{eq:com_to_llf_2}:
\begin{eqnarray}
\fl \textbf{V}_1' = \textbf{V}_{cm} +   \sqrt{| \textbf{V}_1 - \textbf{V}_{cm} |^2 - \frac{2 m_2}{m^2_1 + m_1 m_2} \varepsilon_k} \; \hat{\textbf{u}}_1' \\
\fl \textbf{V}_2' = \textbf{V}_{cm} +   \sqrt{| \textbf{V}_2 - \textbf{V}_{cm} |^2 - \frac{2 m_1}{m^2_2 + m_1 m_2} \varepsilon_k} \; \hat{\textbf{u}}_2'
\end{eqnarray}
where $\hat{\textbf{u}}_1' = \textbf{v}_1' / v_1'$ and $\hat{\textbf{u}}_2' = \textbf{v}_2' / v_2'$ are the unitary velocity vectors. Then we can use \eref{eq:com}:
\begin{eqnarray}
\fl \textbf{V}_1' &= \frac{m_1 \textbf{V}_1 + m_2 \textbf{V}_2}{m_1 + m_2} \cdot \hat{\textbf{U}}_{cm} \\
&+ \frac{m_2}{m_1 + m_2} \sqrt{|\textbf{V}_1 - \textbf{V}_2|^2 - \frac{2 (m_1 + m_2) }{m_1 m_2}  \varepsilon_k} \; \hat{\textbf{u}}_1' \nonumber \\
\fl \textbf{V}_2' &= \frac{m_1 \textbf{V}_1 + m_2 \textbf{V}_2}{m_1 + m_2} \cdot \hat{\textbf{U}}_{cm} \\
&+ \frac{m_1}{m_1 + m_2} \sqrt{|\textbf{V}_2 - \textbf{V}_1|^2 - \frac{2 (m_1 + m_2) }{m_1 m_2}  \varepsilon_k} \; \hat{\textbf{u}}_2' \nonumber
\end{eqnarray}
to calculate particle energies:
\begin{eqnarray}
E_1' = \frac{1}{2} m_1 \left| \frac{m_1 V_1 + m_2 V_2}{m_1 + m_2} \; \hat{\textbf{U}}_{cm} +  \frac{m_2}{m_1 + m_2} \right. \\
\left. \sqrt{V^2_1 + V^2_2 - 2 V_1 V_2 \cos \Theta_2 - \frac{2 (m_1 + m_2) }{m_1 m_2} \varepsilon_k} \; \hat{\textbf{u}}_1' \right|^2 \nonumber \\
E_2' = \frac{1}{2} m_2 \left| \frac{m_1 V_1 + m_2 V_2}{m_1 + m_2} \; \hat{\textbf{U}}_{cm} +  \frac{m_1}{m_1 + m_2} \right. \\
\left. \sqrt{V^2_2 + V^2_1 - 2 V_2 V_1 \cos \Theta_2 - \frac{2 (m_1 + m_2) }{m_1 m_2} \varepsilon_k} \; \hat{\textbf{u}}_2' \right|^2 \nonumber
\end{eqnarray}
where $\hat{\textbf{U}}_{cm}$ is the center of momentum unit vector: 
\begin{eqnarray}
\textbf{V}_{cm} &= \frac{m_1 V_1 (0,0,1) + m_2 V_2 (\sin \Theta_2 , 0, \cos \Theta_2)}{m_1 + m2} \\
&= \frac{m_2 V_2 }{m_1 + m2} \;\; (\sin \Theta_2, 0, \cos \Theta_2 + \frac{m_1 V_1}{m_2 V_2}) \nonumber \\ 
\Gamma &= \sqrt{1 + \left(\frac{m_1 V_1}{m_2 V_2}\right)^2 + 2 \frac{m_1 V_1}{m_2 V_2} \cos \Theta_2} \label{eq:gamma} \\
\hat{\textbf{U}}_{cm} &= \frac{\textbf{V}_{cm}}{V_{cm}} = (\sin \Theta_2, 0, \cos \Theta_2 + \frac{m_1 V_1}{m_2 V_2}) / \Gamma
\end{eqnarray}

Here we can define the quantity:
\begin{eqnarray}
\Delta &= \sqrt{V^2_1 + V^2_2 - 2 V_1 V_2 \cos \Theta_2 - \frac{2 (m_1 + m_2) }{m_1 m_2} \varepsilon_k} \label{eq:delta}
\end{eqnarray}
leading us to the vectorial relation:
\begin{eqnarray}
E_1' &= \frac{1}{2} m_1 \left| \frac{m_1 V_1 + m_2 V_2}{m_1 + m_2} \; \hat{\textbf{U}}_{cm} +  \frac{m_2}{m_1 + m_2} \Delta \; \hat{\textbf{u}}_1' \right|^2 \label{eq:final_e1_vec} \\
E_2' &= \frac{1}{2} m_2 \left| \frac{m_1 V_1 + m_2 V_2}{m_1 + m_2} \; \hat{\textbf{U}}_{cm} +  \frac{m_1}{m_1 + m_2} \Delta \; \hat{\textbf{u}}_2' \right|^2 \label{eq:final_e2_vec}
\end{eqnarray}
or, the scalar form:
\begin{eqnarray}
E_1' = \frac{1}{2} m_1 & \left[ \left(\frac{m_1 V_1 + m_2 V_2}{m_1 + m_2} \right)^2 + \left(\frac{m_2 \Delta}{m_1 + m_2}\right)^2 \right. \label{eq:final_e1_sca}\\
& \qquad + \left. \frac{2 m_2 \Delta \Omega_1 (m_1 V_1 + m_2 V_2)}{\left(m_1 + m_2\right)^2}\right] \nonumber \\
E_2' = \frac{1}{2} m_2 & \left[ \left(\frac{m_1 V_1 + m_2 V_2}{m_1 + m_2} \right)^2 + \left(\frac{m_1 \Delta}{m_1 + m_2}\right)^2 \right. \label{eq:final_e2_sca}\\
& \qquad + \left. \frac{2 m_1 \Delta \Omega_2 (m_1 V_1 + m_2 V_2)}{\left(m_1 + m_2\right)^2}\right] \nonumber
\end{eqnarray} 
given we define the cosine between $\hat{\textbf{U}}_{cm}$ and $\hat{\textbf{u}}'$ as:
\begin{eqnarray}
& \Omega_1 = \hat{\textbf{U}}_{cm} \cdot \hat{\textbf{u}}_1 = \\
& \left( \sin \Theta_2 \sin \theta'_1 \cos \phi_1' + \cos \Theta_2 \cos \theta_1' + \frac{m_1 V_1}{m_2 V_2} \cos \theta_1' \right) / \Gamma \nonumber \\
& \Omega_2 = \hat{\textbf{U}}_{cm} \cdot \hat{\textbf{u}}_2 = \\
& \left( \sin \Theta_2 \sin \theta'_2 \cos \phi_2' + \cos \Theta_2 \cos \theta_2' + \frac{m_1 V_1}{m_2 V_2} \cos \theta_2' \right) / \Gamma \nonumber
\end{eqnarray}


\section{Specific case: $V_2 = 0$ and $\varepsilon_k = 0$}
The simplest verification is to check if we get back to \eref{eq:delta_e_ela} for elastic collisions ($\varepsilon_k = 0$) where the bullet energy is greater than the thermal gas energy ($V_1 \gg V_2$). In this case, we get a finite value for \eref{eq:delta}
\begin{eqnarray}
\Delta &= V_1 
\end{eqnarray}
and an infinite value for \eref{eq:gamma}:
\begin{eqnarray}
\Gamma \rightarrow \infty 
\end{eqnarray}

However, we can calculate the limits getting:
\begin{eqnarray}
\lim\limits_{V_2 \rightarrow 0} \Omega_1 = \cos \theta_1' \\
\lim\limits_{V_2 \rightarrow 0} \Omega_2 = \cos \theta_2'
\end{eqnarray}
allowing reducing \eref{eq:final_e1_sca} and \eref{eq:final_e2_sca} to:
\begin{eqnarray}
E_1' = \frac{1}{2} m_1 & \left[\frac{m^2_1 + m^2_2 + 2 m_1 m_2 \cos \theta_1'}{\left(m_1 + m_2\right)^2}\right] V^2_1 \\
E_2' = \frac{1}{2} m_2 & \left[\frac{2 m^2_1 ( 1 + \cos \theta_2')}{\left(m_1 + m_2\right)^2} \right]  V^2_1 
\end{eqnarray} 
and finally, defining $\Delta E = E - E'$, to:
\begin{eqnarray}
\Delta E_1' = &\frac{ 2 m_1 m_2 (1 - \cos \theta_1')}{\left(m_1 + m_2\right)^2} E_1 \\
\Delta E_2' = - &\frac{2 m_1 m_2 ( 1 + \cos \theta_2')}{\left(m_1 + m_2\right)^2}   E_1
\end{eqnarray} 

\section{Specific case: $V_2 = 0$ and $\varepsilon_k \ne 0$}
For inelastic collisions ($\varepsilon_k \ne 0$), we can replace in \eref{eq:final_e1_sca} and \eref{eq:final_e2_sca} the values $V_1 = \sqrt{2 E_1 / m_1}$ and $V_2 = 0$. In this way we get:
\begin{eqnarray}
\Delta = \sqrt{\frac{2 m_2 E_1 - 2 (m_1 + m_2) \varepsilon_k}{m_1 m_2}} \\
\lim\limits_{V_2 \rightarrow 0} \Omega_1 = \cos \theta_1' \\  
\lim\limits_{V_2 \rightarrow 0} \Omega_2 = \cos \theta_2'  
\end{eqnarray}
and finally:
\begin{eqnarray}
E_1' = &\left[1 - \frac{m_2}{m_1 + m_2} \frac{\varepsilon_k}{E_1} + \right. \\
& \left.  \frac{2 m_1 m_2}{(m_1 + m_2)^2} \left(\sqrt{1 - \frac{m_1 + m_2}{m_2} \frac{\varepsilon_k}{E_1}} \cos \theta_1' - 1 \right)\right] E_1\nonumber \\
E_2' = & \frac{2 m_1 m_2}{(m_1 + m_2)^2} \left[1 - \frac{m_1 + m_2}{2 m_2} \frac{\varepsilon_k}{E_1} + \right. \\
& \left. \sqrt{1 - \frac{m_1 + m_2}{m_2} \frac{\varepsilon_k}{E_1}} \cos \theta_2' \right] E_1  \nonumber
\end{eqnarray}

\section{Plots}
In this section, we present the Monte-Carlo simulation of the drift of electrons in a uniform static electric field. To perform the simulation, we used a framework we developed \cite{renda2020betaboltz,betaboltz,drifter}. A total of \num{25} particles were put in an infinite volume under the effect of a static uniform electric field between \SIrange{1e-3}{1e-1}{\townsend}.

The collision time were calculated using the null-collision technique and the \textit{null-collision} technique \cite{skullerud1968stochastic} while, for the collision kinematics, we used the relations provided by Okhrimovskyy \cite{okhrimovskyy2002electron}. The simulation if stopped when reaching \num{250000} real collisions and, the whole event, is repeated \num{250} times for each field value.

First we performed a simulation at \SI{0}{\kelvin} temperature, where all the gas components are at \textit{rest} and does not have any kinetic energy. Then we repeated the simulation at \SI{20}{\celsius} and at \SI{2000}{\celsius}, to confirm the behavior for increasing temperatures.

\begin{figure*}
	\centering
	\begin{subfigure}{.46\textwidth}
		\includegraphics[width=\linewidth]{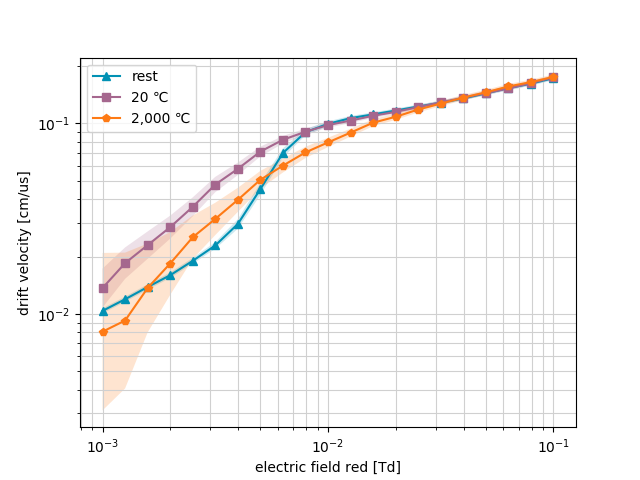}
		\caption{$Ar$}
		\label{fig:plot_velocity_Ar}
	\end{subfigure}
	\begin{subfigure}{.46\textwidth}
		\includegraphics[width=\linewidth]{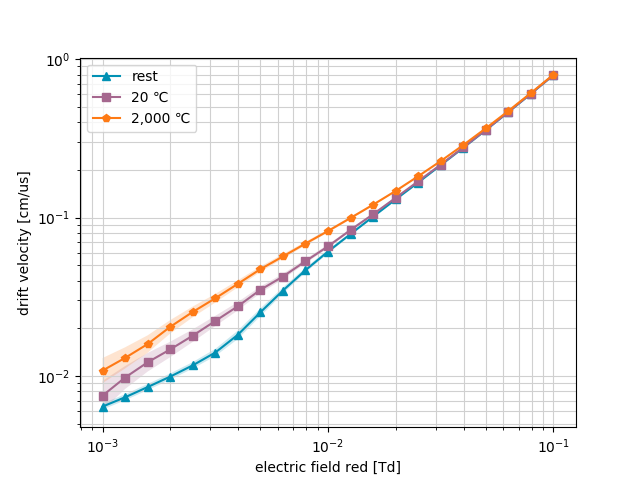}
		\caption{$CF_4$}
		\label{fig:plot_velocity_CF4}
	\end{subfigure}
	\caption{Drift velocities calculated for \num{250} events with \num{25} particles having \num{250000} real collisions each. Cross-section data from \cite{biagi2019lxcat,bsr2020lxcat} for $Ar$ and \cite{bordage2019lxcat} for $CF_4$. Bands represent standard deviations, scaled down by a factor of \num{10} for better graphical representation.}
	\label{fig:plot_velocity}
\end{figure*}

In \fref{fig:plot_velocity}, we can see the simulated drift velocities for a mono-atomic molecule and for a poly-atomic gas with spherical symmetry. We can notice that for higher electric fields, the drift velocities tends to converge. This is expected, because in this region, the mean electron energy is bigger than the thermal energy of the gas components.

\begin{figure*}
	\centering
	\begin{subfigure}{.46\textwidth}
		\includegraphics[width=\linewidth]{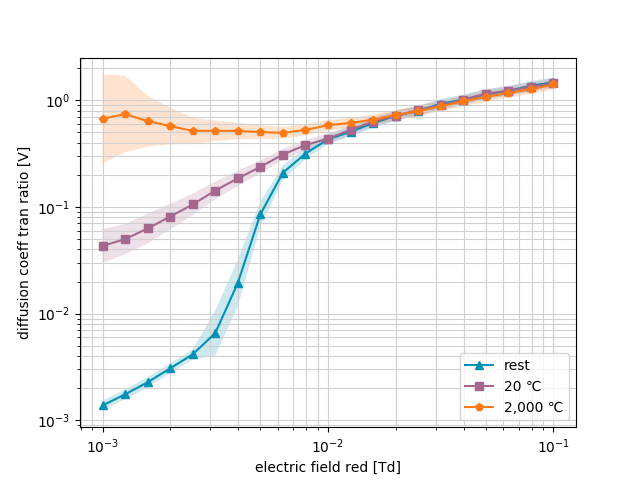}
		\caption{$Ar$}
		\label{fig:plot_diffusion_tran_Ar}
	\end{subfigure}
	\begin{subfigure}{.46\textwidth}
		\includegraphics[width=\linewidth]{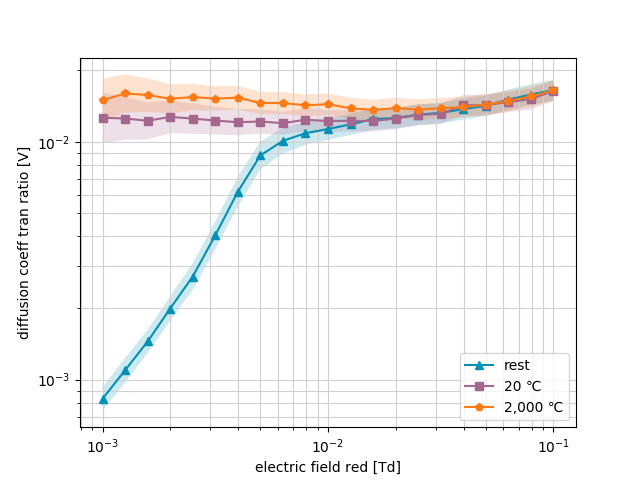}
		\caption{$CF_4$}
		\label{fig:plot_diffusion_tran_CF4}
	\end{subfigure}
	\begin{subfigure}{.46\textwidth}
		\includegraphics[width=\linewidth]{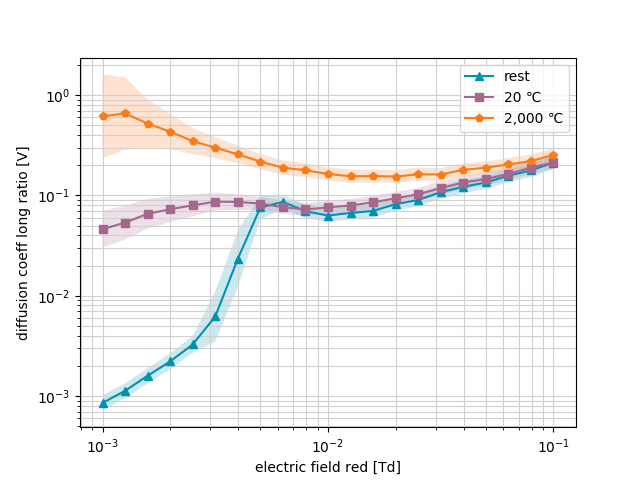}
		\caption{$Ar$}
		\label{fig:plot_diffusion_long_Ar}
	\end{subfigure}
	\begin{subfigure}{.46\textwidth}
		\includegraphics[width=\linewidth]{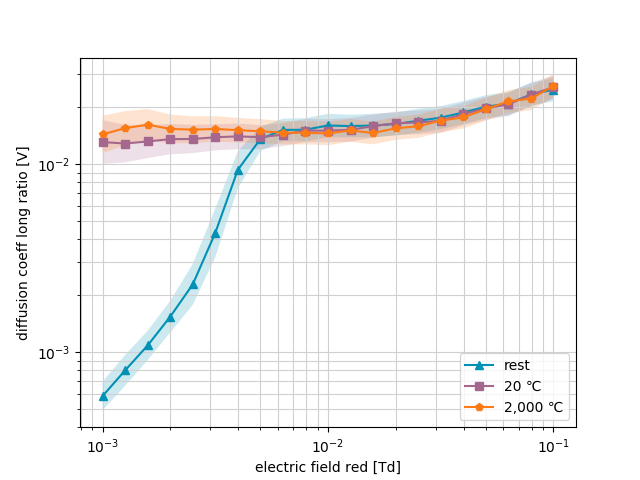}
		\caption{$CF_4$}
		\label{fig:plot_diffusion_long_CF4}
	\end{subfigure}
	\caption{Traversal and longitudinal diffusion coefficient calculated for \num{250} events with \num{25} particles having \num{250000} real-collisions each. Cross-section data from \cite{biagi2019lxcat,bsr2020lxcat} for $Ar$ and \cite{bordage2019lxcat} for $CF_4$. Bands represent standard deviations, scaled down by a factor of \num{10} for better graphical representation.}
	\label{fig:plot_diffusion}
\end{figure*}

In \fref{fig:plot_diffusion}, we can see how, for lower fields, the gas temperature has a direct impact on the diffusion coefficients: the electrons are spread out by the chaotic thermal movement of the gas molecules.

\section{Conclusions}
In this article, we presented our derivation of kinematic formulas to describe electron or ion collisions in gaseous medium at low electric fields, where gas molecules can not be considered at rest. We consider these relations to be useful when performing Monte-Carlo simulation at low $E/N$ values. 

\section*{CRediT author statement}

\textbf{Michele Renda}: Conceptualization, Methodology, Software, Writing - Original Draft \\
\textbf{Iulia Stefania Trandafir}: Writing-Reviewing and Editing,  Formal analysis, Validation

\ack
This work was supported by the research grants \texttt{ATLAS CERN-RO} and \texttt{PN19060104}.

\section*{References}
\bibliographystyle{iopart-num}
\bibliography{bibliography}

\end{document}